\documentclass[journal,12pt,onecolumn,letterpaper]{IEEEtran}
\usepackage{arxiv}
\usepackage{geometry}
\usepackage{times}
\usepackage{cite}
\usepackage{url}
\usepackage{graphicx}
\usepackage{lscape}
\usepackage{subfigure}
\usepackage{rotating}
\usepackage{rotfloat}
\usepackage{xcolor}
\usepackage{amsmath}
\usepackage{amssymb}
\usepackage[linesnumbered,ruled,vlined]{algorithm2e}
\usepackage{pseudocode}
\usepackage{tabularx}
\usepackage{caption}
\usepackage{array}
\usepackage[english]{babel}
\usepackage{gensymb}
\usepackage{textcomp}
\usepackage{placeins}
\usepackage{balance}
\usepackage{booktabs}

\title{MERP: Metaverse Extended Reality Portal\\
}

\author{
Anisha Ghosh\\
Centre of Excellence, Artificial Intelligence \& Robotics (AIR),\\
School of Computer Science and Engineering\\
VIT-AP University, India \\
\texttt{ghoshanisha2002@gmail.com}\\
\And
Aditya Mitra \\
Centre of Excellence, Artificial Intelligence \& Robotics (AIR),\\
 School of Computer Science and Engineering\\
 VIT-AP University, India \\
\texttt{adityamitra5102@gmail.com}\\
\And
Anik Saha\\
Centre of Excellence, Artificial Intelligence \& Robotics (AIR),\\
 School of Computer Science and Engineering\\
 VIT-AP University, India \\
\texttt{adityamitra5102@gmail.com}\\
\And
Sibi Chakkaravarthy Sethuraman \\
Centre of Excellence, Artificial Intelligence \& Robotics (AIR),\\
School of Computer Science and Engineering\\
VIT-AP University, India \\
\texttt{sb.sibi@gmail.com} \\
\And
Anitha Subramanian \\
Centre of Excellence, Artificial Intelligence \& Robotics (AIR),\\
School of Electronics Engineering\\
VIT-AP University, India \\
\texttt{anithachubbu@gmail.com} \\
}

\begin{document}

\maketitle

\begin{abstract}
A standardized control system called Metaverse Extended Reality Portal (MERP) is presented as a solution to the issues with conventional VR eyewear. The MERP system improves user awareness of the physical world while offering an immersive 3D view of the metaverse by using a shoulder-mounted projector to display a Heads-Up Display (HUD) in a designated Metaverse Experience Room. To provide natural and secure interaction inside the metaverse, a compass module and gyroscope integration enable accurate mapping of real-world motions to avatar actions. Through user tests and research, the MERP system shows that it may reduce mishaps brought on by poor spatial awareness, offering an improved metaverse experience and laying the groundwork for future developments in virtual reality technology.
MERP, which is compared with existing Virtual Reality (VR) glasses used to traverse the metaverse, is projected to become a seamless, novel and better alternative. Existing VR headsets and AR glasses have well-known drawbacks that making them ineffective for prolonged usage as it causes harm to the eyes.
\end{abstract}


\section{Introduction}

As VR technology gains traction, concerns about user safety and accidents resulting from a lack of awareness of the actual surroundings have heightened. In this paper, we present the MERP, a standardized control system that allays these worries by radically altering the metaverse user experience.
The MERP system, in contrast to traditional VR glasses, uses a shoulder-mounted projector to transmit a Heads-Up Display (HUD) into a specified Metaverse Experience Room. In addition to providing a fully immersive 3D picture of the metaverse, this method allows users to keep spatial awareness of their surroundings, lowering the danger of accidents. The MERP system improves user safety and engagement by bridging the gap between the real world and the metaverse.
The MERP system does this by including a gyroscope and compass module, which precisely record the user's motions in the real world and smoothly translate them to actions taken by their avatars in the metaverse. This easy-to-use interface improves user engagement and immersion while lowering the danger of mishaps that come with conventional VR equipment. This allows for a truly immersive experience, as the user's avatar moves in the virtual world just as they do in the real world. Imagine being able to walk around in the virtual world and have your avatar move with you in real-time - it's like being able to enter the virtual world and truly become a part of it.
But the benefits of the MERP don't stop there. It can support all AR/VR/ER/XR applications with little or no modification, making it extremely versatile and user-friendly. This means that users can experience a wide range of virtual reality applications, from games to educational programs, without having to purchase separate headsets for each application.

\textit{The contributions in this paper include}

\begin{itemize}

\item Cutting-edge control system that tracks user's physical motions and maps them to metaverse avatar activities
\item Shoulder-mounted projector displays a Heads-Up Display (HUD) in a dedicated Metaverse Experience Room
\item Improved user awareness of surroundings and immersive 3D view of the metaverse
\item Safer and more immersive experience compared to traditional VR glasses
\item Reduces accidents through enhanced spatial awareness and collision avoidance
\item Versatile option for gaming, education, telepresence, and other fields

\end{itemize}

\section{Literary Survey}

Virtual reality has become increasingly popular in recent years, with many people turning to VR headsets to experience games, educational programs, and even virtual tours of far-off places. Wiederhold et al. [7] utilized Virtual Reality to mobilize health care. It was tested for anxiety and pain treatments  [7]. The movement modalities have been studied with respect to the virtual reality to help developers design more accessible and seamless virtual reality models [8].  With proper usage and implementation, studies say that Virtual Reality could be a disruptive technology, just like the smartphone and internet as it makes communication and activities seamless even from a large distance [9]. It has further given way to the concept of Digital Twins where robots or other hardware can be controlled from the virtual reality at a distance. This has been instrumental in the development of robotic games with consumer robots [10].  It has been very useful and effective in the education sector and promises better immersive learning experience [11]. Another research has shown the use of Gesture based input devices for the virtual reality which are usually very expensive and are only suitable for scientific and advanced technological experiments \cite{Cnf2}. They have not been adopted yet for day-to-day gaming, entertainment or other purposes. But while VR headsets have come a long way in recent years for different applications and usecases, they still have their limitations. 

One major limitation of VR headsets is that they rely on VR glasses to display the virtual world. While these glasses have certainly come a long way in terms of image quality and comfort, they still expose the user to harmful radiations. Prolonged exposure to these radiations can lead to serious health issues, which is a major concern for many users. The eye is the main health issue. As per Martin Banks these VR glasses has numerous possible problems and it affects the growth of the eye which can lead to myopia or near-sightedness \cite{Jrl1}. Another major issues are Headaches and dizziness which have become more common in those who have used VR because it can disrupt the eye-brain connection. Many games bring objects too quickly to the eye and produce eye strain.

Another limitation of VR headsets is that they cut the user off from their immediate surroundings. While this can certainly enhance the sense of immersion in the virtual world, it can also be dangerous. There have been numerous reports of people injuring themselves or damaging property while using VR headsets. For example, Taylor Murray, who first cut himself playing virtual boxing and then broke a vase weeks later playing virtual tennis, as reported to the Wall Street Journal. Another incident includes 12-year-old Landon Woodward of Rockwood, Mich., who slammed his hand onto a desk while playing a VR game, resulting in a nail falling off. James McLay, 30, of Falkirk, Scotland, bought a mat to stand on so he can feel the area where it is safe for him to play in VR after an incident in which he swung his hand with full force into a shelf while playing a sword-fighting game. These incidents highlight the danger of using VR headsets that cut the user off from their surroundings \cite{Onl1}.

Another study on Metaverse based learning is proposed by Gim et al. \cite{cnf2}. Their study explored variables that affect learner satisfaction, mediating the flow theory \cite{Cnf1}. Their Results showed that the most variables of the self-determination theory and information system quality within VR education impact learner satisfaction. However, students with myopia or hypermetropia suffered and this resulted in increase in the number of eye diseases and migraines. Also, due to extremely high price of good quality VR sets, many were forced to compromise on the quality. 

Metaverse is targeted to have hyper-spatio-temporality, meaning a digital world having both space and time. Users can interact with one another in real time as well as have a sensation of the place where they are \cite{Jrl2}. It can be a seamless meeting space for humans in the form of avatars as well as bots and it would be difficult, if not impossible to differentiate between them. Metaverse can be used to play, work and socialize \cite{Jrl3}. However, it is still a challenge to deal with the health hazards and safety issues.

With MERP, all these problems have been tacked as it required users to just be comfortable in their usual environment. It is a low-cost, shoulder-mounted projector that projects the virtual world onto a screen in front of the user. This means that the user is not required to wear VR glasses, which protects them from harmful radiations. It also means that the user is not cut off from their surroundings, making it much safer to use. Table \ref{TBL:Comparison} shows the differences between traditional VR Glasses and MERP. 

\begin{table*}[h!]
\centering
\caption{Differences between traditional VR glasses and MERP.}
\label{TBL:Comparison}

\begin{tabular}{@{}lll@{}}
\toprule
Aspect & Traditional VR Glasses & MERP \\ \midrule
Health hazards to eyes, risk of myopia.  & Yes   & No. \\
Health hazards for radiation & Yes   & No.  \\
User awareness of the real world. & No & Yes  \\
Bulky to wear in front of the eye & Yes  & No  \\
\begin{tabular}[c]{@{}l@{}}Integrable with any game, \\ metaverse or other application\end{tabular} & Usually no & Yes, with key-bindings \\ \bottomrule
\end{tabular}

\end{table*}

\section{Proposed Work}

This paper proposes a cutting-edge control system that tracks a user's physical motions and maps them to the activities taken by their avatar in the metaverse. In a dedicated Metaverse Experience Room, a shoulder-mounted projector shows a Heads-Up Display (HUD), which improves user awareness of the surrounding surroundings while offering an immersive 3D picture of the metaverse. In comparison to conventional VR glasses, MERP delivers a safer and more immersive experience by reducing accidents through enhanced spatial awareness and collision avoidance. It is a flexible option for engaging in virtual worlds because of its potential uses in gaming, education, telepresence, and other fields.

\subsection{Architecture}

MERP employs sensors to record the movements of the user. A compass is used to find the differential movement of the user about the magnetic north pole of the earth. This is then mapped into mouse movement. Figure 1 shows the user movement in the polar axes.

\begin{figure}[h!]
\vspace*{-10pt}
\centerline{\includegraphics[width=17.5pc]{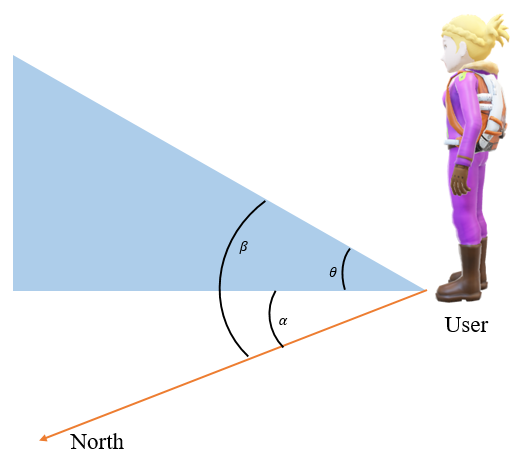}}
\caption{User movement in polar axes.}
\end{figure}

If $\alpha$ is the initial angle made by the user with true north as recorded by the compass module, and $\beta$ is the angle made by the user after moving, then the resultant angle moved by the user is. 

\begin{equation}
\Theta = \beta - \alpha 
\end{equation}

\begin{figure}[h!]
\vspace*{-10pt}
\centerline{\includegraphics[width=17.5pc]{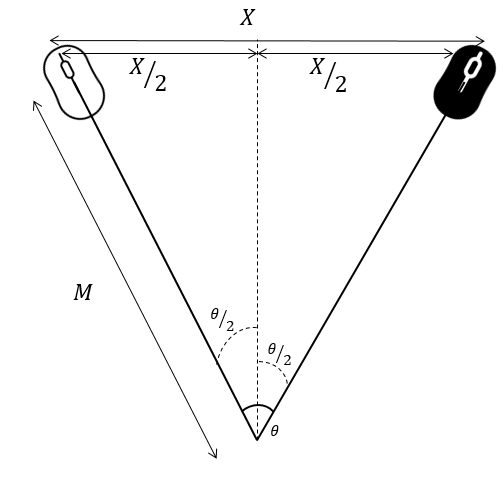}}
\caption{Intended mouse movement.}
\label{FIG:Mouse-Movement}
\end{figure}

This is further translated into the movement of the mouse. There is a calibration factor M which is directly proportional to how sensitive the mouse is, or the mouse speed as adjusted in the system settings. The mouse movement in horizontal direction, X can be calculated with $\theta$ and M. It is to be noted that the unit for X is pixels. Figure. \ref{FIG:Mouse-Movement} shows the intended mouse movement.

\begin{equation}
\frac{X}{2} = M \ast \sin \sin \left ( \frac{\Theta}{2} \right )
\end{equation}

\begin{equation}
 \frac{X}{2} = M \ast \sqrt{\frac{1 - \cos \cos \Theta }{2}}
\end{equation}

\begin{equation}
X = 2M \ast \sqrt{\frac{1 - \cos \cos \Theta }{2}}
\end{equation}

\begin{figure}[htbp]
\vspace*{-10pt}
\centering
\includegraphics[width=17.5pc]{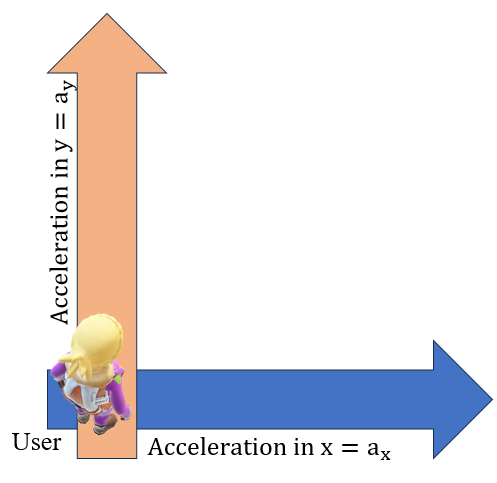}
\caption{User movement in cartesian axes}
\label{FIG:User_Movement-Cartesian}
\end{figure}

The movement of the user on cartesian axes is calculated using an accelerometer and then translated into keyboard input. Figure. \ref{FIG:User_Movement-Cartesian} shows the user movement in cartesian axes. The acceleration of the user in x and y directions are $a_x$ and $a_y$.

\begin{equation}
     \begin{array}{l}
Velocity\ in\ x:v_{x} \ =\ \int a_{x} dt\\
\\
Distance\ moved\ in\ x:d_{x} \ =\int v_{x} dt\ \\ =\ \iint a_{x} dt\ *\ dt
\end{array}
\end{equation}

\begin{equation}
     \begin{array}{l}
Velocity\ in\ y:v_{y} \ =\ \int a_{y} dt\\
\\
Distance\ moved\ in\ y:d_{y} \ =\int v_{x} dt\ \\ =\ \iint a_{y} dt\ *\ dt
\end{array}
\end{equation}







The integration factor dt can be interpreted as the time needed for one CPU cycle of the microcontroller.

\begin{equation}
dt = \frac{1}{Clock speed of MCU}
\end{equation}

A Keyboard calibration factor K is introduced. This signifies for how much time the keyboard keys are to be emulated. If $d_x$ is positive, the right movement key is emulated for $K*d_{x}$ time. Similarly, if $d_{x}$ is negative, the left movement key is emulated for $K*d_{x}$ time. If $d_{y}$is positive, the forward movement key is emulated for $K*d_{y}$ time. Similarly, if $d_{y}$ is negative, the backward movement key is emulated for $K*d_{y}$ time.

The Human Interface Device (HID) inputs was emulated to a metaverse environment using a microcontroller. It was then projected using a shoulder mounted projector in a VR experience room.

\subsection{Experimental Setup and Results}

\begin{figure}[h!]
\vspace*{-10pt}
\centering
\includegraphics[width=17.5pc]{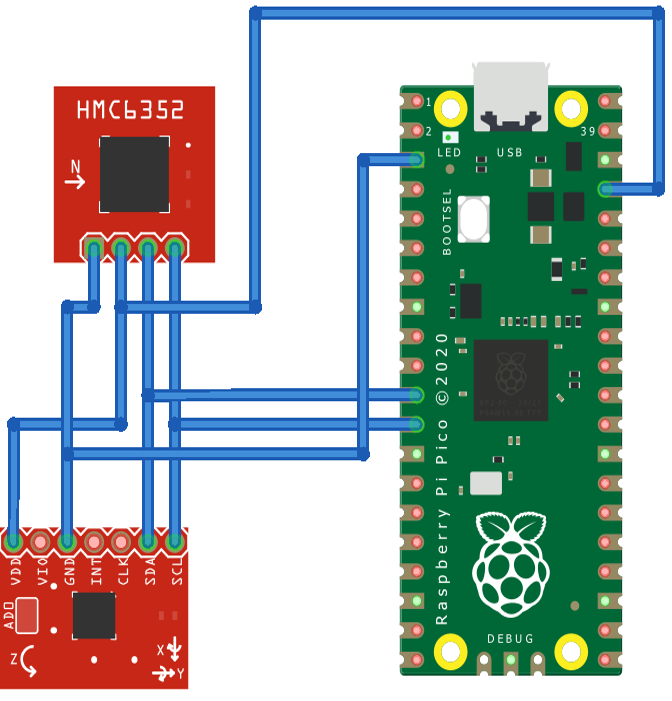}
\caption{Experimental setup used for MERP}
\label{FIG:Exp_Setup}
\end{figure}

\begin{figure*} [htbp]
\centering
\subfigure[Avatar selection in metaverse]{
	\centering
	\includegraphics[width=0.8\textwidth]{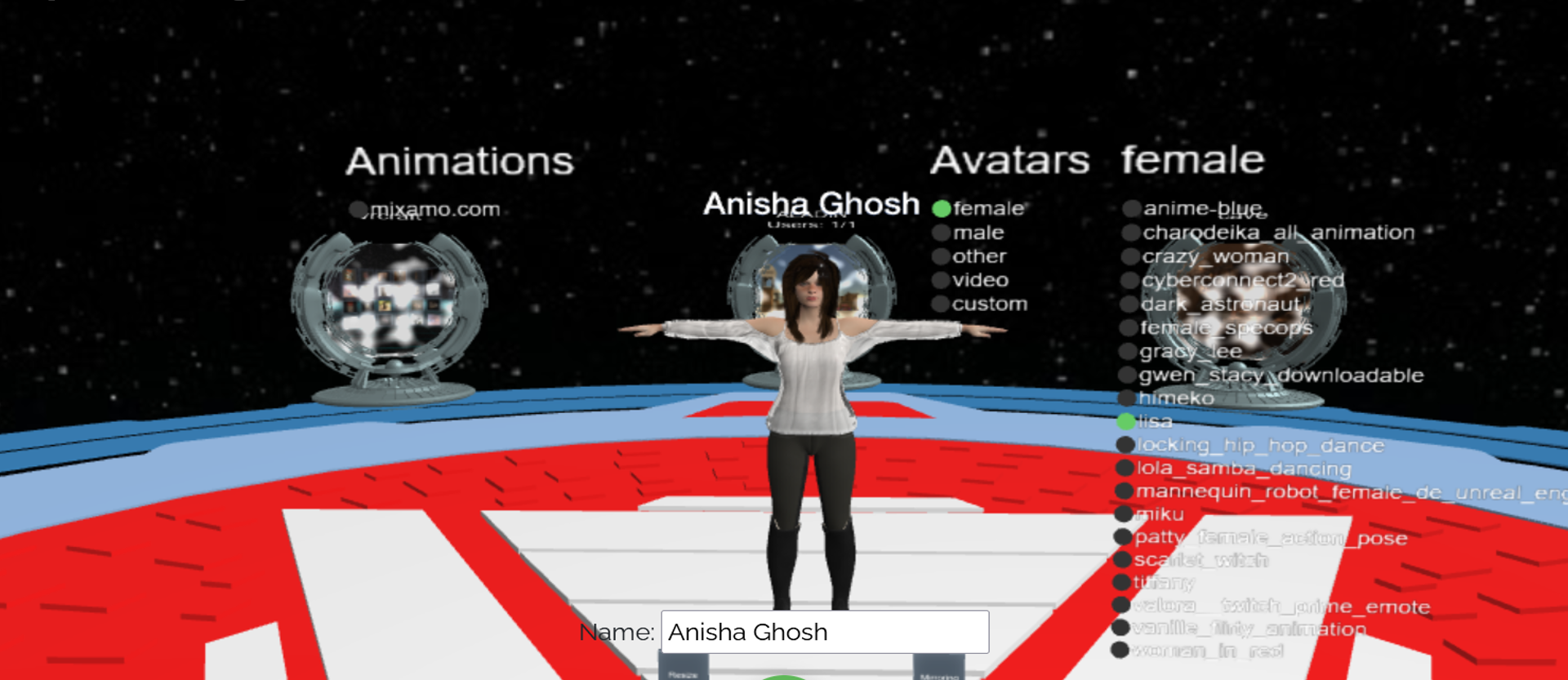}
	 \label{FIG:Avatar-Selection} 
}
\subfigure[Avatar traversing the metaverse in First Person View using MERP]{
	\includegraphics[width=0.8\textwidth]{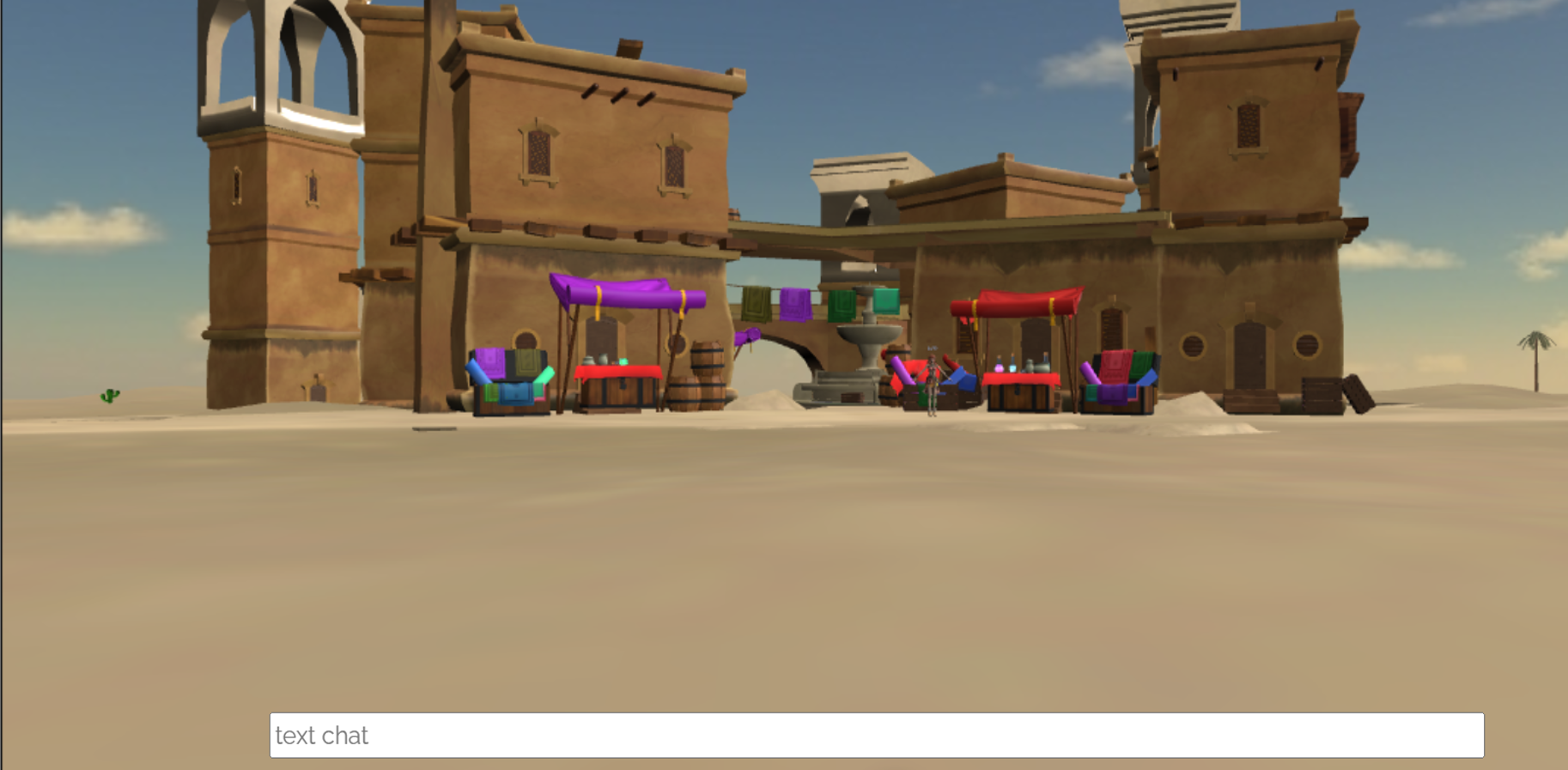}
	\label{FIG:Avatar-Movement}
}
\caption{Experimental Results of MERP}
\label{FIG:Exp-Result}
\end{figure*}

Figure. \ref{FIG:Exp_Setup} shows the experimental setup of MERP. A HMC6352 Magnetic compass sensor and MPU6050 3-axis gyroscope was employed to sense the user’s movements in the polar and cartesian coordinates respectively. They were connected by the Inter-Integrated Circuit (I2C) Bus with the Data (SDA) and Clock (SCL) pins. The compass module and the gyroscope could communicate with the RPi Pico microcontroller parallelly. It was then encoded into the Keystrokes and mouse movements using a Raspberry Pi Pico containing the RP2040 microcontroller with a clock speed of 133 MHz It then emulated the keyboard and mouse using the Adafruit HID library on CircuitPython framework. It was tested on a clone of VRSpace.org which is an open source metaverse developed using Babylon.js. The results were projected using a ViewSonic M1 Mini Portable projector on a VR experience room. This gave the desired results with zero to minor latency between the user’s movements and the avatar’s movements.

\subsection{Experimental Results}
MERP was able to map the user’s movements in the real world to the Avatar’s movements in the virtual world. MERP was tested with five volunteers who communicated with each other in the metaverse using MERP. Their reports indicated that MERP worked as desired with nil to low latency between the user’s movements and the avatar’s movements. 
Figure. \ref{FIG:Avatar-Selection} shows the avatar selection in metaverse and Figure. \ref{FIG:Avatar-Movement} shows traversing the metaverse with MERP.

\section{Use Cases}

MERP can be used in the following ways:

\begin{itemize}
\item Gaming: The MERP technology enables players to physically move and interact inside virtual game environments, transforming the gaming experience. The seamless conversion of real-world movements into in-game activities improves immersion and realism whether playing through challenging battle sequences, adventuring across huge fantasy regions, or solving puzzles.
\item Education: The MERP system has a lot of promise for use in educational settings. Students can go on virtual field excursions where they can have a greater feeling of presence while learning about historical locations or natural wonders. To provide a secure and engaging learning environment, training programmes can use the technology to imitate real-life situations like medical procedures or dangerous surroundings.

\item Telepresence and Remote Collaboration: The MERP system offers improved telepresence encounters that let users experience being present even when they are in distant locations. Colleagues, friends, or family may connect in-depth and realistically by projecting a virtual image of each user into a communal virtual area, bypassing the restrictions of physical distance.

\item Fitness and exercise: By using the MERP system in fitness apps, exercising may be made fun and immersive. Exercise regimens may be made more pleasurable and motivational by allowing users to engage in virtual workouts while measuring their motions and receiving real-time feedback.

\item Social Virtual Reality: By enabling users to physically move and make gestures that mimic real-life social signs and expressions, the MERP system improves social interactions in virtual reality. Virtual get-togethers and meetings improve in their immersiveness and engagement, promoting a sense of presence and connection among attendees.

\item Virtual tourism: Using the MERP system, users may explore digital recreations of actual locales, enabling them to virtually travel to famous landmarks and important cultural sites. Without the necessity for physical travel, this may give a distinctive travel experience by giving a window into many locales and cultures.
\end{itemize}

\section{Conclusion}

The MERP (Metaverse Extended Reality Portal) system, which addresses user safety issues and improves user immersion in virtual reality experiences, marks a significant advancement in metaverse control devices. The MERP system enables users to stay aware of their surroundings while exploring the metaverse by utilising a shoulder-mounted projector and a Heads-Up Display (HUD) projected onto a real-world Metaverse Experience Room. Compass and gyroscope integration provides precise translation of physical motions to avatar actions, enabling natural and fluid engagement in the virtual world.

It has been proven via intensive research and user surveys that the MERP system successfully reduces accidents brought on by a lack of awareness of the physical surroundings, giving users a safer and more immersive metaverse experience. The MERP system changes how we interact with virtual reality by bridging the gap between the actual and virtual worlds. Applications in gaming, education, telepresence, and other areas are made possible.

\section{Future Scope}

The MERP system sets the groundwork for future developments in virtual reality and metaverse control devices. Additional sensors and technologies may be incorporated in the future through research and development to improve user interactions in the metaverse. For instance, integrating haptic feedback systems can provide consumers a more tactile experience by giving virtual objects and situations a sensation of touch. This can also be modified to observe obstacles using visualization [12]. 

The MERP system also creates opportunities for teamwork and social engagement within the metaverse. The MERP system can encourage new kinds of shared virtual experiences and enable meaningful interactions by allowing several users to synchronise their avatars and communicate in real-time. We conclude that a signing avatar is a promising way to provide hearing people access to the world of the deaf. MERP can be modified for deaf persons can wear glasses to track the translation with the speaker's facial expressions, gestures, and body language[13].





\begin{thebibliography}{1}
\bibitem{Jrl1}
Koulieris, G.A., Bui, B., Banks, M.S. and Drettakis, G., ``Accommodation and comfort in head-mounted displays,`` {\it ACM Transactions on Graphics (TOG)}., vol. 36, no. 4, pp. 1-11, 2017. (journal)

\bibitem{Onl1}
Noor Al-Sibai, “Virtual Reality users keep suffering horrible injuries,“ [Online]. Available: {https://futurism.com/neoscope/vr-injuries} (URL)

\bibitem{Cnf1}
Gim, G., Bae, H. and Kang, S., “Metaverse Learning: The Relationship among Quality of VR-Based Education, Self-Determination, and Learner Satisfaction,“ {\it In 2022 IEEE/ACIS 7th International Conference on Big Data, Cloud Computing, and Data Science (BCD)}, pp. 279-284, August. 2022. (conference proceedings) 

\bibitem{Cnf2}
Anthes, C., García-Hernández, R. J., Wiedemann, M., and Kranzlmüller, D., "State of the art of virtual reality technology." {\it In 2016 IEEE aerospace conference}, pp. 1-19, IEEE, 2016. (conference proceedings)

\bibitem{Jrl2}
Ning, H., Wang, H., Lin, Y., Wang, W., Dhelim, S., Farha, F., Ding, J. and Daneshmand, M.,  "A Survey on the Metaverse: The State-of-the-Art, Technologies, Applications, and Challenges", {\it IEEE Internet of Things Journal}., 2023. (journal)

\bibitem{Jrl3}
Wang, Y., Su, Z., Zhang, N., Xing, R., Liu, D., Luan, T.H. and Shen, X., "A survey on metaverse: Fundamentals, security, and privacy", {\it IEEE Communications Surveys \& Tutorials}., 2022. (journal)

\bibitem{Jrl4}
Wiederhold, B.K., Miller, I.T. and Wiederhold, M.D., "Using virtual reality to mobilize health care: Mobile virtual reality technology for attenuation of anxiety and pain", {\it IEEE Consumer Electronics Magazine}, vol. 7, no. 1, pp.106-109, 2017. (journal)

\bibitem{Jrl5}
Ap Cenydd, L. and Headleand, C.J., "Movement modalities in virtual reality: a case study from ocean rift examining the best practices in accessibility, comfort, and immersion", {\it IEEE Consumer Electronics Magazine}, vol. 8, no. 1, pp.30-35, 2018. (journal)

\bibitem{Jrl6}
Rosedale, P., "Virtual reality: The next disruptor: A new kind of worldwide communication", {\it IEEE Consumer Electronics Magazine}, vol. 6, no. 1, pp.48-50, 2016. (journal)

\bibitem{Jrl7}
Prattico, F.G. and Lamberti, F., "Mixed-reality robotic games: design guidelines for effective entertainment with consumer robots", {\it IEEE Consumer Electronics Magazine}, vol .10, no. 1, pp.6-16, 2020. (journal)

\bibitem{Jrl8}
Pagano, K., Haddad, A. and Crosby, T., "Virtual reality-making good on the promise of immersive learning: The effectiveness of in-person training, with the logistical and cost-effective benefits of computer-based systems", {\it IEEE Consumer Electronics Magazine}, vol. 6, no. 1, pp.45-47, 2016. (journal)


\bibitem{Cnf3}
Maruta, K., Takizawa, M., Fukatsu, R., Wang, Y., Li, Z. and Sakaguchi, K., "Blind-spot visualization via AR glasses using millimeter-wave V2X for safe driving." {\it In 2021 IEEE 94th Vehicular Technology Conference (VTC2021-Fall)}, pp. 1-5, IEEE, 2021. (conference proceedings)

\bibitem{Cnf4}
Nguyen, L.T., Schicktanz, F., Stankowski, A. and Avramidis, E., "Evaluating the translation of speech to virtually-performed sign language on AR glasses." {\it In 2021 13th International Conference on Quality of Multimedia Experience (QoMEX)}, pp. 141-144, IEEE, 2021. (conference proceedings)

\end{thebibliography}
\end{document}